\documentclass[prl,superscriptaddress,showpacs,longbibliography,reprint]{revtex4-1}

\usepackage{hyperref}

\usepackage{color}
\usepackage[usenames,dvipsnames]{xcolor}
\usepackage{amsmath,amsthm,amssymb}
\usepackage{graphicx}
\usepackage{epsfig}
\usepackage{dcolumn}
\usepackage{bm}
\usepackage{mathrsfs}
\usepackage{multirow}
\usepackage[all]{xy}
\usepackage{pbox}
\usepackage{verbatim}
\usepackage{braket}
\usepackage{dsfont}

\DeclareMathOperator{\Tr}{Tr}
\def\(({\left(}
\def\)){\right)}
\def\[[{\left[}
\def\]]{\right]}

\newcommand{\be}{\begin{equation}}
\newcommand{\ee}{\end{equation}}
\newcommand{\ben}{\begin{eqnarray}}
\newcommand{\een}{\end{eqnarray}}
\newcommand{\beq}{\begin{equation}}
\newcommand{\eeq}{\end{equation}}

\usepackage{graphicx}

\usepackage{color,soul}

\newcommand{\ketbra}[2]{\ket{#1}\hspace{-2.1pt}\bra{#2}}

\begin{document}

\title{Numerical Simulation of Critical Quantum Dynamics without Finite Size Effects}

\author{Edward Gillman}
\affiliation{School of Physics and Astronomy, University of Nottingham, Nottingham, NG7 2RD, UK}
\affiliation{Centre for the Mathematics and Theoretical Physics of Quantum Non-Equilibrium Systems,
University of Nottingham, Nottingham, NG7 2RD, UK}

\author{Federico Carollo}
\affiliation{Institut f\"{u}r Theoretische Physik, Universit\"{a}t T\"{u}bingen, Auf der Morgenstelle 14, 72076 T\"{u}bingen, Germany}

\author{Igor Lesanovsky}
\affiliation{School of Physics and Astronomy, University of Nottingham, Nottingham, NG7 2RD, UK}
\affiliation{Centre for the Mathematics and Theoretical Physics of Quantum Non-Equilibrium Systems,
University of Nottingham, Nottingham, NG7 2RD, UK}
\affiliation{Institut f\"{u}r Theoretische Physik, Universit\"{a}t T\"{u}bingen, Auf der Morgenstelle 14, 72076 T\"{u}bingen, Germany}

\begin{abstract}
Classical $(1+1)D$ cellular automata, as for instance Domany-Kinzel cellular automata, are paradigmatic systems for the study of non-equilibrium phenomena. Such systems evolve in discrete time-steps, and are thus free of time-discretisation errors. Moreover, information about critical phenomena can be obtained by simulating the evolution of an initial seed that, at any finite time, has support only on a finite light-cone. This allows for essentially numerically exact simulations, free of finite-size errors or boundary effects. Here, we show how similar advantages can be gained in the quantum regime: The many-body critical dynamics occurring in $(1+1)D$ quantum cellular automata with an absorbing state can be studied directly on an infinite lattice when starting from seed initial conditions. This can be achieved efficiently by simulating the dynamics of an associated one-dimensional, non-unitary quantum cellular automaton using tensor networks. We apply our method to a model introduced recently and find accurate values for universal exponents, suggesting that this approach can be a powerful tool for precisely classifying non-equilibrium universal physics in quantum systems.
\end{abstract}

\date{\today}

\maketitle

\textit{Introduction.}--- One of the most intriguing aspects of non-equilibrium phase transitions (NEPTs) and of many-body critical dynamics is the emergence of universal behaviour: systems with very different microscopic details can display the same scaling laws at a macroscopic scale, both for key stationary and dynamical quantities such as correlations or order parameters. As in equilibrium, an understanding of such critical features comes from their classification into universality classes \cite{Hinrichsen2000, Lubeck2005, Henkel2008}. Each class groups systems with the same emergent behaviour, according to parameters known as critical exponents. However, in contrast to equilibrium settings, even the simplest critical non-equilibrium systems, e.g.~those featuring absorbing state phase transitions in the directed percolation (DP) universality class, are not analytically solvable and their exponents cannot be determined exactly.

\begin{figure*}[t]
\centering
\includegraphics[width=1\linewidth]{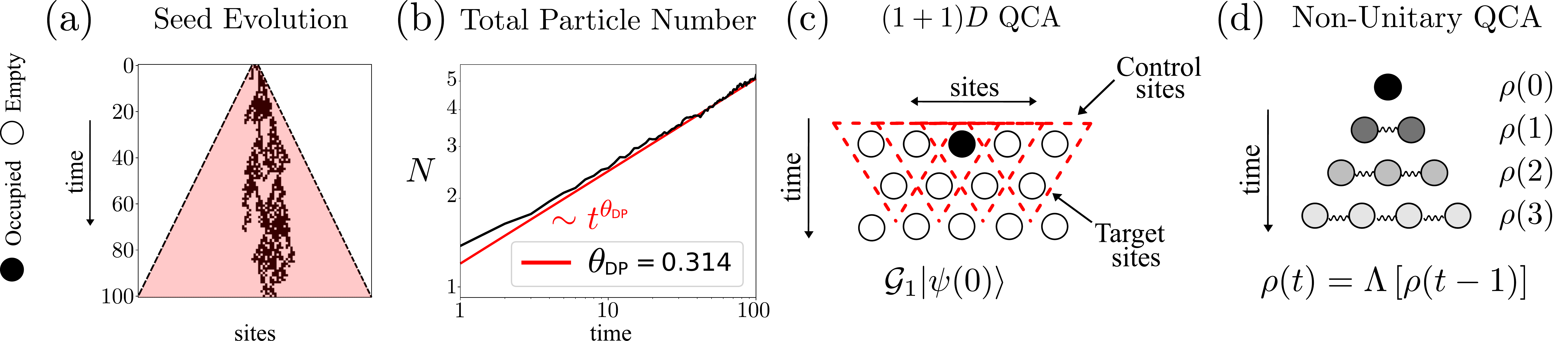}
\caption{\textbf{Seed evolutions in a $\mathbf{(1+1)D}$ (quantum) cellular automaton:} \textbf{(a)} Seed evolution of the classical Domany-Kinzel cellular automaton at the critical site-DP point \cite{Henkel2008} performed directly on an infinite lattice. Occupied sites only fall inside the indicated light-cone (dashed lines and shaded region). \textbf{(b)} The total number of occupied sites, $N(t)$, is shown averaged over $1000$ runs (solid black line). Even for relatively short times the universal power-law can be seen (straight solid red line), and the observed exponent is in agreement with the expected value for $1D$ DP: $\theta_{\text{DP}} = 0.314$. \textbf{(c)} In a $(1+1)D$ QCA, a $2D$ lattice is initiated in a product state with all empty sites apart from the first row which encodes the initial condition. The state $\ket{\psi(1)}$, obtained by updating the target sites in the subsequent row via the application of three-body unitary gates, is shown. The operator $\mathcal{G}_1$ is the product of the applied gates, thus $\ket{\psi(1)}=\mathcal{G}_1\ket{\psi(0)}$. \textbf{(d)} When $(1+1)D$ QCA on an infinite lattice display a strict light cone, the reduced dynamics of a row with seed initial conditions are fully captured by those of a finite-size reduced state, $\rho(t)$. The size of this state grows proportionally with time. This allows for the dynamics of the underlying unitary $(1+1)D$ QCA to be studied via the corresponding non-unitary dynamics of $\rho(t) = \Lambda[\rho(t-1)]$, free of finite-size effects.}
\label{fig:Seed_QCA_Schematic}
\end{figure*}

To overcome this problem, efficient numerical schemes for simulating non-equilibrium many-body dynamics are required. This concerns both continuous time models, such as the contact process \cite{Grassberger1979}, and discrete time evolutions like in the paradigmatic Domany-Kinzel cellular automata (DKCA) \cite{PhysRevLett.53.311}. For analyzing these classical systems, it is convenient to study critical behavior following a local perturbation of the absorbing state \cite{Grassberger1979}. For instance, in the DKCA this is a state with a single occupied site, as shown in Fig. \ref{fig:Seed_QCA_Schematic}(a). The importance of such single-seed scenarios is two-fold. Firstly, numerically-exact simulations can be performed directly in the limit of an infinite system, i.e.~free of finite-size effects. This stems from the fact that the information about the presence of a local perturbation propagates with a strict light cone. Secondly, as a consequence, critical exponents can be extracted directly from such evolutions by considering only a finite portion of the system, see Fig. \ref{fig:Seed_QCA_Schematic}(b).

For many-body quantum systems, tracking the evolution of an initial seed also provides access to key universal quantities. However, the corresponding simulations in continuous \cite{Gillman2019, Minjae2020} and discrete time \cite{Gillman2020,Wintermantel2020} pose significant challenges. Indeed, unlike their classical counterparts, quantum models with absorbing states display in general superposition and entanglement \cite{Marcuzzi2016,Buchhold2017,Roscher2018,Carollo2019}, which make numerical simulations demanding.

In this paper, we introduce a method to study critical quantum non-equilibrium behaviour that builds on the advantages of seed simulations. We show that the discrete-time dynamics of $(1+1)D$ quantum cellular automata (QCA) starting from a single seed can be simulated without finite-size effects, by using a tensor network (TN) \cite{Schollwock2011,PerezGarcia2007,
Crosswhite2008,Pirvu2010,Paeckel2019,Jarkovsky2020,
Verstraete2004,Eisert2013,Montangero2018,Orus2019,Ran2020} that grows dynamically. Just like their classical counterparts --- which include the DKCA ---  $(1+1)D$ QCA are free of time-discretization errors. As such, our approach offers an extremely clean, flexible and efficient framework for studying NEPTs in quantum systems. To demonstrate its potential, we apply it to previously studied QCA \cite{Gillman2020}. The method introduced here allows for the accurate estimation of critical exponents at significantly reduced computational costs.

\textit{Single-seed dynamics in QCA.}--- Similarly to the case of classical cellular automata \cite{RevModPhys.55.601} [c.f. Fig. \ref{fig:Seed_QCA_Schematic}(a)], the full information about $(1+1)D$ QCA is encoded in a two-dimensional (tilted) lattice, as shown in Fig. \ref{fig:Seed_QCA_Schematic}(c). The horizontal dimension represents space, while the vertical one provides a notion of time \cite{Lesanovsky2019,Gillman2020,PhysRevE.100.020103,Wintermantel2020}. Each lattice site is described by a two-level system, with basis states $\ket{\bullet},\ket{\circ}$ denoting an occupied or an empty site, respectively. The lattice is initialized with all sites in the empty state, except for those in the zeroth row, which encode the initial condition. 

The evolution of this $2D$ lattice occurs via the action of unitary operators (gates) on lattice sites. These gates act on pairs of consecutive rows, such that at time-step $t$, the ``target" row $t$, is updated according to the state of ``control" row $t-1$. For concreteness, we consider here a local update rule for the $(1+1)D$ QCA based on three-body gates, $G_{t,k}$, but our findings can be generalized to other scenarios. The gate $G_{t,k}$ performs a controlled unitary operation on the target site at $(t,k)$, with controls at $(t-1,k_{-})$ and $(t-1,k_{+})$, where $k_{-}$ ($k_{+}$) refers to the control site to the left (right) of target site $k$. In order for the QCA to feature an absorbing state, we impose a constraint on $G_{t,k}$: we assume that target sites are not modified whenever the corresponding control sites are both found in the empty state \cite{Gillman2020}. As such, if a control row has all sites empty,  no update takes place on their targets.
 
The global update for the entire row, $\mathcal{G}_{t}$, is then an ordered product of the gates $G_{t,k}$, one per target site. In contrast to classical systems, one must pay special attention to the ordering of the unitary quantum gates, as these do not commute in general. As such, to preserve a physical notion of causality --- a concept which is also key to the definition of QCA in the field of quantum information (QI) \cite{Farrelly2019,Piroli2020} --- only those gate orderings giving rise to a strict light cone will be considered. We remark that, while a specific ordering will affect the exact values of the observables, the underlying universal physics of the model is not expected to change from one ordering to the next.

Due to the unitarity of the gates, the state of the $2D$ lattice after $t$ time steps is pure, $\ket{\psi(t)}$. It contains the full space-time information of the QCA and can be used to compute unequal time observables, such as time-correlation functions. However, here we focus on observables which can be computed from the reduced state of the QCA on row $t$ at time $t$. These observables provide sufficient information to determine the universality class of the considered model \cite{Grassberger1979,Henkel2008,Gillman2020}.

Mathematically, the reduced state is given by $\varrho(t)=\Tr'\left(\ket{\psi(t)}\bra{\psi(t)}\right)$, where $\Tr'$ is the partial trace over all sites with the exception of those in row $t$. The evolution of $\varrho(t)$ describes the discrete-time dynamics of a $1D$ system. Corresponding to the classical case where irreversible $1D$ CA can be simulated by reversible $(1+1)D$ CA \cite{Toffoli1990}, the dynamics of $\varrho(t)$ are in general non-unitary. Since gates act solely on consecutive rows, the evolution of $\varrho(t)$ can be defined iteratively as
\begin{align}
\varrho(t) = \Tr_{t-1}\left[ \mathcal{G}_{t} \varrho(t-1) \otimes \ket{\Omega_t}\bra{\Omega_t}\mathcal{G}_{t}^{\dagger} \right]\,,
\label{eqn:rdm_update_rule}
\end{align}
where $\ket{\Omega_t}$ is the $t$-th row configuration with all empty sites and $\Tr_{t-1}$ indicates the trace over the sites where $\varrho(t-1)$ is defined.

Turning now to dynamics ensuing from a single-seed initial condition, we set $\varrho(0)$, i.e.~the zeroth row of the $2D$ lattice, to be $\varrho(0)=\sigma^{+}_{\text{seed}} \ket{\Omega_{0}}\bra{\Omega_{0}}\sigma^{-}_{\text{seed}}$, where $\sigma^{+} = \ketbra{\bullet}{\circ}$ and $\sigma^-=(\sigma^+)^\dagger$, in such a way that the seed site (at the centre of the initial row) is occupied. Clearly, for this choice of the initial state, $\varrho(0)$ factorises as $\varrho(0)=\rho_{\circ}\otimes\rho(0)\otimes \rho_{\circ}$, where the matrix $\rho_{\circ}$ indicates an infinite tensor product of empty states, while $\rho(0) = \ketbra{\bullet}{\bullet}$ has support only on a single site. 

The most striking consequence of the existence of a strict light cone is that, for dynamics starting from a state with finite non-trivial support such as the single-seed, at any time $\varrho(t)$ can be factorised as $\varrho(t)=\rho_{\circ}\otimes\rho(t)\otimes \rho_{\circ}$. Here, $\rho(t)$ has support only on a finite set of sites, $\mathcal{L}_t$, with size $L_{t} = |\mathcal{L}_{t}|$. Consequently, the entire reduced dynamics of the $(1+1)D$ QCA can be captured without finite size effects through the evolution of $\rho(t) = \Lambda\left[\rho(t-1)\right]$, where $\Lambda$ is the map that implements this update, see Fig. \ref{fig:Seed_QCA_Schematic}(d).

In general, starting from any $\rho(t-1)$ and for any gate ordering, the reduced dynamics can be implemented via Eq. \eqref{eqn:rdm_update_rule} as follows. First $\rho(t-1)$ is mapped into $\varrho(t-1)$ by attaching an infinite product of empty sites in row $t-1$ to the left and right of $\mathcal{L}_{t}$. Second, row $t$ is included in a product state of all empty sites. Third, the gates are applied via $\mathcal{G}_{t}$ before, finally, the sites of row $t-1$ are traced out.

\begin{figure}[t]
\centering
\includegraphics[width=1\linewidth]{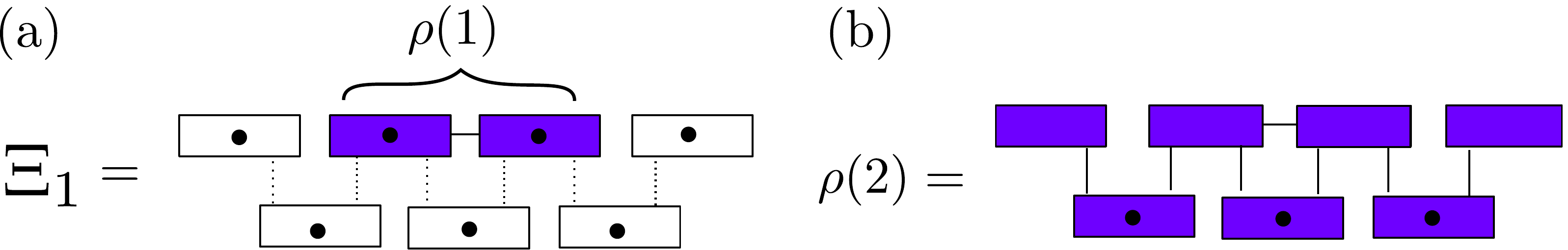}
\caption{\textbf{Reduced dynamics of the $\mathbf{(1+1)D}$ QCA:} \textbf{(a)} To evolve $\rho(t-1) \to \rho(t)$ through the map $\Lambda$, we begin via an MPO representation of $\rho(t-1)$, shown here for $t=2$. Empty sites are then added at locations where gates act non-trivially. This operation  defines the state $\Xi_{t-1}$. We have depicted these states here using the standard diagrammatic notation for TNs \cite{Schollwock2011, Montangero2018}. In this notation, tensors are represented by shapes with a number of legs corresponding to their order. Each tensor corresponding to operators in a local Hilbert space must have two legs for the ``physical" indices, here represented by a solid black circle. Other ``virtual" legs which join the shapes (solid black lines) encode correlations between these, and we denote the trivial legs (indicating no correlations) as dashed lines. \textbf{(b)} The gates are then applied in MPO form to update the state. By tracing out the sites of row $t-1$, indicated diagrammatically by removing physical legs, the exact TN representation of $\rho(t)$ is obtained. Approximating this by an MPO  allows the scheme to be iterated.}
\label{fig:MPO_evo}
\end{figure}

For orderings with strict light cones, this procedure simplifies since  sites outside the support $\mathcal{L}_t$ are in the absorbing state and, thus, only a finite number of gates in $\mathcal{G}_t$ act non-trivially. Therefore, the map $\Lambda$ can be implemented by considering only a finite number of extra empty sites and gates (see also Fig.~\ref{fig:MPO_evo}). In addition, for any given $t$, $L_{t} \le L_{t-1} + v$ for some fixed integer $v$ determined by the gate and the ordering. In what follows, we show how $\Lambda$ can be expressed  in terms of a finite TN that updates a matrix product operator (MPO) representation for $\rho(t-1)$ into an MPO representation of $\rho(t)$. This enables efficient numerical simulations of the dynamics of $\rho(t)$, allowing us to investigate universal aspects of the QCA, directly in the infinite lattice limit.

\textit{TNs for seed evolutions on infinite lattices.}--- For the sake of clarity, we now specify a gate ordering. We choose an alternating leftmost-rightmost ordering, where first the leftmost target site is updated, then rightmost, then the second leftmost and so on. Generalization to other gate orderings is possible (see Supplemental Material \cite{SM} for a discussion of another example). The alternating leftmost-rightmost ordering leads to the lowest possible increase in $L_{t}$, i.e. $v=1$, and thus has minimal computational cost.

It is convenient to represent $\rho(t)$ as an MPO. The map $\Lambda$, which connects two MPOs with different supports, can then be understood in terms of a TN, see Fig.  \ref{fig:MPO_evo}. At any given time, $\rho(t-1)$ is represented as an MPO [c.f. Fig. \ref{fig:MPO_evo}(a)] with maximum bond-dimension $\chi$. To find the representation for $\rho(t)$, with our choice of the ordering, we first expand $\rho(t-1)$ by introducing a single empty site at both boundaries and $t+1$ empty sites (the target sites) representing the subsequent row. This defines a new state $\Xi_{t-1}$, with the same non-trivial part. At this point, we can apply all the gates acting non-trivially on the QCA, as shown in Fig. \ref{fig:MPO_evo}(b). This is achieved by representing gates as three-site MPOs and applying these to the previous TN for $\Xi_{t-1}$. To obtain a TN for $\rho(t)$, we then trace out all sites related to row $t-1$. An exact representation of $\rho(t)$ as an MPO can be achieved by factorising the tensors in row $t-1$ and contracting them into those of row $t$. However, such an operation will, in general, lead to an exponential growth of the bond-dimension $\chi$ with time, making numerical simulations infeasible.  To avoid this, the last step of the update consists in constructing an approximate MPO for $\rho(t)$. There are several  strategies for approximating $\rho(t)$ using an MPO with fixed $\chi$. For TNs, a natural approach is to first map the MPO into a matrix product state (MPS), apply standard approximation methods available for MPS \cite{Schollwock2011, Paeckel2019}, and, finally, map the MPS back into an MPO \cite{SM}.

\begin{figure*}[t]
\centering
\includegraphics[width=0.9\linewidth]{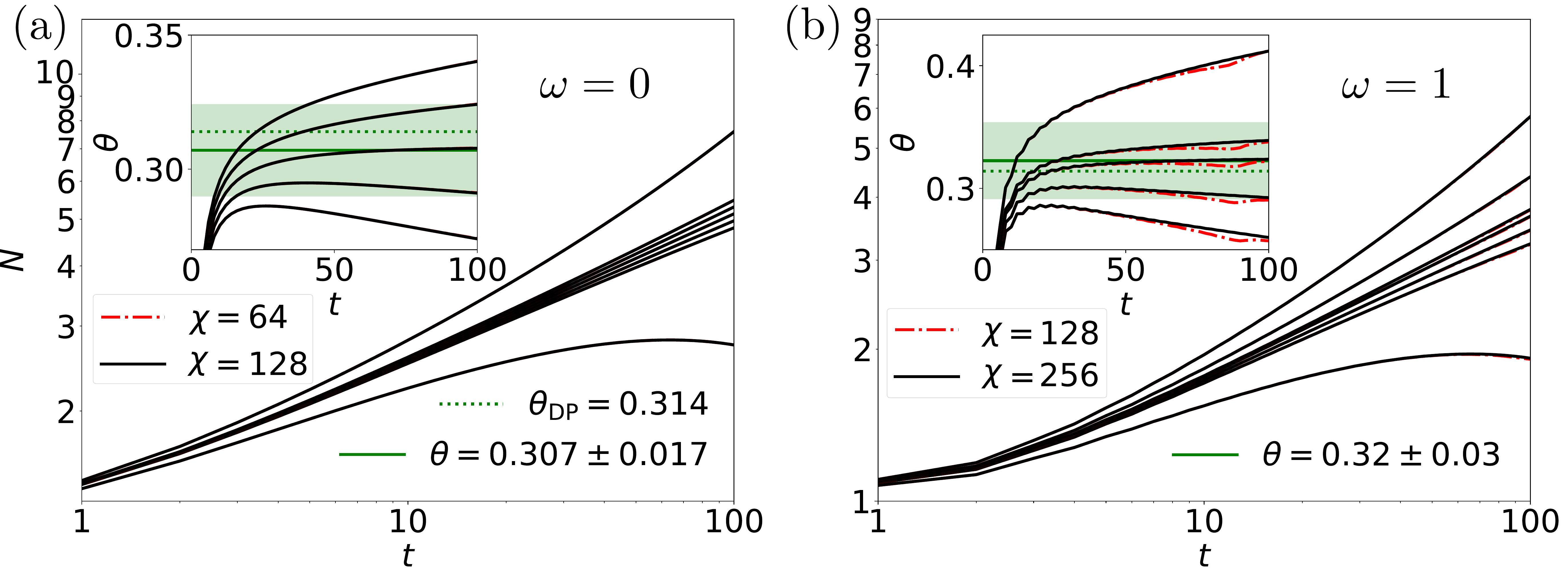}
\caption{\textbf{Critical behavior:} \textbf{(a)} The evolution of the total number of occupied sites, $N(t)$, is shown for $t \in \left[1, 100\right]$ for various values of $\Gamma$. This includes five intermediate values close to the critical point --- as indicated by the almost linear behaviour in the log-log plot --- and two values (the bottom-most and top-most lines) further from the critical point, illustrating the NEPT from a state of zero particles to one with a diverging number. In ascending order from the lowest line, the values are  $\Gamma$ are $0.98, 0.995, 0.996, 0.997, 0.998, 0.999, 1.01$. The inset shows $\theta(t)$ for the central five values. From these, the flattest curve provides the estimate for the critical point, $\Gamma_{c} = 0.997 \pm 0.01$, and exponent, $\theta = 0.307 \pm 0.017$. The shaded region represents the error \cite{SM}. The estimated value of $\theta$ is consistent with that of $1D$ DP, indicated by the dotted green line. $\chi$ is the maximum MPO bond-dimension used. \textbf{(b)} Corresponding plots for $\omega = 1$ with $\Gamma = 1.015, 1.03, 1.032, 1.034, 1.035, 1.04, 1.05$. These produce the estimates $\Gamma_{c} = 1.034 \pm 0.02$ and $\theta = 0.32 \pm 0.03$, also consistent with $1D$ DP.}
\label{fig:Total_particles}
\end{figure*}

\textit{Critical Exponents for $(1+1)D$ QCA.}---To demonstrate the potential of the method introduced here, we consider the $(1+1)D$ QCA defined by the local gate,
\begin{align}
G_{t,k} =\exp\left[-i \Gamma \left( U_{t-1,(k_{-},k_{+})} P_{t-1,(k_{-},k_{+})}  \sigma^{+}_{t,k} + \text{h.c.}\right)\right]\, 
\label{eqn:gate_def}
\end{align}
This gate implements a generalised rotation of the target by an angle $\Gamma$, conditioned on the controls by a two-body projector,
\begin{align}
P_{t,(k_{-},k_{+})}  = \mathds{1}_{t,(k_{-},k_{+})} - \ketbra{\circ\circ}{\circ\circ}_{t,(k_{-},k_{+})}.
\end{align}
This dynamics thus has the absorbing state $\ket{...\circ\circ...}$ which follows from $P_{t,(k_{-},k_{+})}\ket{\circ\circ}_{t,(k_{-},k_{+})}=0$ \cite{Lesanovsky2019}. In order to control the degree of quantum correlations \cite{Gillman2020}, we introduce the two-body unitary
\begin{align}
U_{t,(k_{-},k_{+})} = \exp\left(-i\omega\left[\sigma^{z}_{t,k_{-}} \sigma^{y}_{t,k_{+}} +\sigma^{y}_{t,k_{-}} \sigma^{z}_{t,k_{+}}\right]\right) ~ ,
\label{eqn:unitary}
\end{align}
where $\sigma^{y} = -i \ketbra{\bullet}{\circ} + i \ketbra{\circ}{\bullet}, \sigma^{z} = \ketbra{\bullet}{\bullet} - \ketbra{\circ}{\circ}$. When $\omega = 0$ no entanglement is created in $\varrho(t)$ and it is always separable. As $\omega$ is increased $\varrho(t)$ can become entangled, before again becoming separable when $\omega = \pi/2$.

This particular $(1+1)D$ QCA was studied previously for $\omega = 0$ \cite{Lesanovsky2019} and  with $\omega > 0$ \cite{Gillman2020}. In the first case, the separability of $\varrho(t) $ allowed for the universality class to be established as $1D$ DP, via a mapping to the site-DP critical point of the DKCA. For $\omega > 0$, TNs were used to find bounds on the critical exponent $\alpha$, associated to the decay of the average particle density when starting from homogeneous --- all sites occupied --- initial conditions \cite{Henkel2008}. These exponents were also found to be consistent with $1D$ DP. However, the accuracy of estimates were limited by the computational difficulty of the simulation. This was found to depend strongly on the value of $\omega$, and cases where $U_{t,(k_{-},k_{+})}$ generated significant entanglement were particularly challenging. As such, simulations with values such as $\omega = 1$ led to rather loose bounds on the estimate of $\alpha$.

Here, we study the dynamics of this $(1+1)D$ QCA  starting from an initial seed state and using an alternating leftmost-rightmost gate ordering. To test our method, we consider the challenging $\omega = 1$ case, using $\omega = 0$ for comparison. We focus on the total number of occupied sites at time $t$,
\begin{align}
N(t) &= \sum_{k \in \mathcal{L}_{t}}\Tr\left[\hat{n}_{k}\rho(t)\right] \, ,
\label{eqn:power_laws}
\end{align}
where $\hat{n}_{k}$ is the operator $\hat{n}=\ketbra{\bullet}{\bullet}$ at a given site. At the critical $\Gamma$, this average value is expected to display a universal power-law behavior with critical exponent $\theta$, $N(t)\sim t^\theta$ [see Fig. \ref{fig:Seed_QCA_Schematic}(c)]. We can thus use $N(t)$ both to determine the critical point $\Gamma_{c}$ for each $\omega$ and to estimate the value of $\theta$, as shown in Fig. \ref{fig:Total_particles}. 

For fixed $\omega$, we take several values of $\Gamma$ and simulate $\rho(t)$ up to $t = 100$ for different $\chi$, the two highest of which are shown in Fig.~\ref{fig:Total_particles}. We then calculate the effective exponent, $\theta(t) = \log_{2}\left[N(t)/N(t/2)\right]$, which converges to a constant for power-law behaviour and provides and approximation for the exponent $\theta$ at criticality. Using the highest value of $\chi$ available and taking the curve for which $\theta(t)$ is closest to a constant, we estimate the critical value $\Gamma_{c}$ as well as the exponent $\theta$. Errors due to finite $\chi$ are estimated via the difference of curves with alternative $\chi$ values. For errors associated to the estimation of the critical point, values of $\theta$ extracted from curves with $\Gamma \approx \Gamma_{c}$ are used. For both $\omega = 0$ and $\omega = 1$, the errors due to the estimate of $\Gamma_{c}$ are far larger than those attributable to finite $\chi$. As such, the errors stated in Fig. \ref{fig:Total_particles} correspond to those induced by the estimate of the critical point \cite{SM}.

For $\omega = 0$ and $\omega = 1$, the estimated values of $\theta$ were $\theta = 0.307 \pm 0.017$ and $\theta = 0.32 \pm 0.03$ respectively. Both are consistent with $1D$ DP. Since the errors on these estimates are dominated by the resolution of the grid used to find $\Gamma_{c}$, they can be reduced easily by finer searches. This is in stark contrast to the homogeneous case. Not only are the overall errors there larger due to the presence finite-size effects, but it is the error due to finite $\chi$ that limits accuracy \cite{Gillman2020}.

\textit{Conclusions and Outlook.}--- We have introduced a general scheme for the simulation of seed evolutions in $(1+1)D$ QCA with an absorbing state. This allows for the study of quantum NEPTs free of finite-size effects. This method can be used to provide  an accurate estimates of the critical exponents related to the universal dynamics of these models. Owing to the universality of continuous NEPTs, the method introduced here can be applied well beyond the particular considerations of QCA to the study of out-of-equilibrium quantum many-body systems with absorbing states more broadly. Furthermore, by considering systems with trivial (infinite temperature) steady states, it can easily be extended to the study of systems without absorbing states as well.

Nonetheless, QCA are also of interest in their own right. Not only can they be considered as computational models and analysed from the perspective of QI \cite{Wiesner2009,Cirac2017,Arrighi2019,Farrelly2019}, but, as they are quantum many-body systems, their emergent physical properties can be intriguing \cite{Hillberry2020}. In this regard, our method, along with $(1+1)D$ QCA more generally, may be rather useful as it allows for the explicit study of emergent behavior in QCA --- including non-unitary QCA, far less studied than their unitary counterparts \cite{Richter1996,Brennen2003,Piroli2020,Wintermantel2020}. Applying these tools for characterizing non-equilibrium universality classes will potentially provide general insights into the relationships between the computational properties of QCA and their collective many-body behavior.

\acknowledgments We acknowledge support from The Leverhulme Trust (Grant No. RPG-2018-181) the “Wissenschaftler-R\"uckkehrprogramm GSO/CZS” of the Carl-Zeiss-Stiftung and the German Scholars Organization e.V., as well as  through  the  Deutsche  Forschungsgemeinsschaft(DFG, German Research Foundation) under Project No.435696605, and under Germany's Excellence Strategy - EXC No. 2064/1 - Project No. 390727645. FC acknowledges support through a Teach@T\"ubingen Fellowship. We are grateful for access to the University of Nottingham’s Augusta HPC service. We also acknowledge the use of Athena at HPC Midlands+, which was funded by the EPSRC on Grant No. EP/P020232/1, in this research, as part of the HPC Midlands+ consortium.

\bibliographystyle{apsrev4-1}
\bibliography{QCA_bib}

\onecolumngrid
\newpage

\pagebreak
\widetext

\begin{center}
\textbf{\large Supplemental Materials}
\end{center}

\setcounter{section}{0}
\setcounter{equation}{0}
\setcounter{figure}{0}
\setcounter{table}{0}
\setcounter{page}{1}
\makeatletter

\renewcommand\thesection{S\arabic{section}}
\renewcommand{\theequation}{S\arabic{equation}}
\renewcommand{\thefigure}{S\arabic{figure}}
\renewcommand{\bibnumfmt}[1]{[S#1]}

\section{Reduced State Dynamics with Odd-Even Scheme}

In this section, we consider the reduced state dynamics $\rho(t) = \Lambda\left[\rho(t-1)\right]$ for the more common ``odd-even" gate ordering. In this case, the update of a given target row $t$ occurs by first updating all odd targets and then all even ones. Since the fundamental gate $G$ acts on two control sites, all odd (even) targets can be updated simultaneously since the corresponding uniraty operators commute.

Defining $\mathcal{G}_{t}$ to be the operator that updates the full row of targets, this ordering defines the decomposition,
\begin{align}
\mathcal{G}_{t} = \mathcal{G}_{t}^{e}\mathcal{G}_{t}^{o} ~,
\end{align}
where,
\begin{align}
\mathcal{G}_{t}^{e} &= \prod_{k ~ \text{even}} G_{t,k} ~, \\
\mathcal{G}_{t}^{o} &= \prod_{k ~ \text{odd}} G_{t,k} ~.
\end{align}

\begin{figure*}[b]
\centering
\includegraphics[width=1\linewidth]{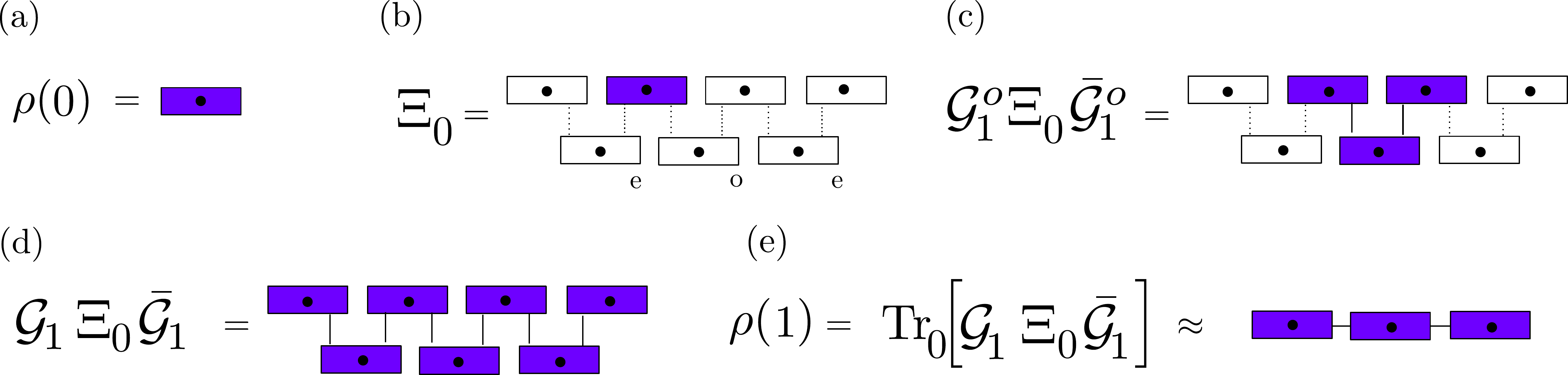}
\caption{\textbf{The reduced dynamics of the $\mathbf{(1+1)D}$ QCA under an odd-even update scheme:} \textbf{(a)} The update begins with the MPO representation of $\rho(t-1)$, choosing $t = 1$ here. \textbf{(b)} Empty sites are then added to form two-row state $\Xi_{t-1}$. In the present scheme, a single site to the left of the current reduced state MPO and two to the right can be used. In the target layer $t+2$ sites which have two of the previous control sites as parents on the tilted square lattice are added. \textbf{(c)} After decomposing $G$ as a three site MPO, the odd update is applied, with the first site in the target row being considered as even for this purpose. Here, the overbar indicates complex conjugation. \textbf{(d)} The even update is then applied, such that all $L_{t} = L_{t-1} + 2 $ targets have been updated. \textbf{(e)} Finally, the control row $t-1$ is traced out, and the resulting two-layer network represents $\rho(t)$. This can subsequently be used to form an approximation of $\rho(t)$ as an MPO.}
\label{fig:MPO_evo_odd_even}
\end{figure*}

This ordering of the gates defines a reduced dynamics $\Lambda$ for which the non-trivial part of the reduced density matrix for a row, $\rho(t)$ increases by two-sites at every time-step. In other words, this scheme givevs rise to a strict light cone structure where $v = 2$. The corresponding update using tensor networks is illustrated in Fig. \ref{fig:MPO_evo_odd_even}, with all details contained in the caption.

\section{Approximation of Reduced Dynamics using Matrix Product Operators}

When dealing with the approximation of pure states using TNs, one often uses matrix product states (MPS) \cite{Schollwock2011,Montangero2018,Paeckel2019}. In this framework, one takes an MPS of fixed bond-dimension, $\ket{\tilde{\phi}}$, and uses it as a variational ansatz to approximate some desired state, $\ket{\phi}$, also represented as an MPS but with higher bond-dimension. This is achieved by adjusting the parameters contained in $\ket{\tilde{\phi}}$ (the elements of the tensors that define it) so as to minimise the Hilbert-space distance between the two states.

To apply this idea to matrix product operators (MPOs) and density matrices, one can map the density matrices to states in the doubled space via the isomorphism $\ketbra{m}{n} \to \ket{m}\ket{n}$ which implements $\rho(t) \to \ket{\rho(t)}$. In terms of the TNs, this maps the MPO representation of $\rho(t)$ to an MPS representation of $\ket{\rho(t)}$ in a straightforward manner by collecting together the physical ``bra and ket" indices of the MPO into a single compound index for each site.

A state $\ket{\tilde{\rho}(t)}$ represented as an MPS can then be used as a variational approximation of $\ket{\rho(t)}$ by solving the minimisation problem,
\begin{align}
\ket{\tilde{\rho}(t)} = \text{argmin}_{\ket{\tilde{\rho}(t)} \in \text{MPS}(\chi)} | \ket{\rho(t)} - \ket{\tilde{\rho}(t)} |^{2} ~,
\end{align}
where $\text{MPS}(\chi)$ indicates the set of MPS states with maximum bond-dimension $\chi$.

Finally the MPS representation of $\ket{\tilde{\rho}(t)}$ can be mapped into an MPO representation by factorising the physical indices that were previously grouped.

In the context of the reduced dynamics $\rho(t) = \Lambda\left[\rho(t-1)\right]$ considered in the main text, given an initial MPO representation of $\rho(t-1)$, we produce an approximate MPO representation of $\rho(t)$ as follows:
\begin{enumerate}
\item Represent the state $\ket{\rho(t)}$ as a two-layer network, by collecting the physical indices of the corresponding representation of $\rho(t)$.
\item  Initiate an MPS ansatz state, $\ket{\tilde{\rho}}$.
\item Iteratively minimise the Hilbert-space norm between these two states, sweeping through site-by-site in the MPS of $\ket{\tilde{\rho}}$ (i.e. sequentially minimise the parameters contained in a tensor corresponding to a particular site while keeping the others fixed) to make efficient use of computational resources. This is achieved using standard MPS methods \cite{Schollwock2011}.
\item Perform sweeps until a chosen observable has converged to sufficient accuracy. In the main text, we use the total number of excitations between the exactly updated $\rho(t)$ and the variational approximation.
\item Map the resulting MPS $\ket{\tilde{\rho}}$ to an MPO, which is then taken as the approximation of $\rho(t)$ for subsequent iterations.
\end{enumerate}

We note that, while this procedure is optimal for pure states, there is no such guarantee for mixed states, where the natural distance measure between states is not the Hilbert-space norm used for the objective function of the minimisation, but instead the trace norm. However, the procedure has proved effective in practice. In any case, we emphasise that this approximation step can be replaced by any other desired method, as it is independent of the overall approach taken in the main text.

\section{Estimation of $\theta$ Exponent from $N(t)$}

In this section, we provide details on the estimation procedure used for the critical point, $\Gamma_{c}$, and the critical exponent $\theta$, which established the values found for the $(1+1)D$ QCA \eqref{eqn:gate_def} considered in the main text and displayed in Fig. \ref{fig:Total_particles}.

To estimate $\Gamma_{c}$, we simulate seed evolutions and calculate $N(t)$ for a grid of $\Gamma$, the resolution of which sets the fundamental error in the estimation of $\Gamma_{c}$.

Constructing the effective exponent, $\theta(t)$, for each value of $\Gamma$, we estimate $\Gamma_{c}$ as the value of $\Gamma$ for which the curve of $\theta(t)$ is closest to a constant. This is measured by the gradient averaged between $t \in \left[50,100\right]$, using the simulation with the highest value of $\chi$. The curve with the lowest absolute averaged gradient is then chosen for the estimate.

The error on this estimate is then taken as the maximum difference between this value of $\Gamma$ and that of the closest values of $\Gamma$ above and below. If these are asymmetrically spaced around $\Gamma_{c}$, the larger value is chosen. As such, the estimate of the critical point can be improved by taking finer grids in $\Gamma$, as is usual in analysis of classical systems \cite{Henkel2008}. In the main text, we estimate that $\Gamma_{c} = 0.997 \pm 0.01$ and $\Gamma_{c} = 1.034 \pm 0.02$ for $\omega = 0$ and $1$ respectively.

The value of $\theta$ is then estimated from the value of $\theta(t)$ for $\Gamma_{c}$, averaged over $t = \left[50, 100\right]$. To estimate the error in this value induced by the error in $\Gamma_{c}$, we take the maximum difference between this value and those calculated similarly for the values of $\Gamma$ directly above and below. This gives $\theta = 0.307 \pm 0.017$ and $\theta = 0.322 \pm 0.031$ for $\omega = 0$ and $\omega = 1$ respectively. We note this is much larger than the errors associated to the finite bond-dimension effects, discussed below, and hence this is the overall error stated in the main text.

To estimate the errors induced by restricting the value of $\chi$ in simulations, we take the absolute difference between the value of the observable at that time, and the value obtained from simulations with $\chi/2$. In the main text, the two highest values of $\chi$ used were $\chi = 128$ and $256$ for $\omega = 0$ and $\omega = 1$ respectively. As such, simulations with $\chi = 64$ and $128$ for $\omega = 0$ and $\omega = 1$ were used to estimate the finite bond-dimension errors.

In the case studied in the main text, the errors in $N(t)$ due to $\chi$ where are most $0.09\%$ and $3.75 \%$ for $\omega = 0, 1$, taken over all values of $t \le 100$ and $\Gamma$. The corresponding errors propagated to $\theta(t)$ can be seen visually in Fig. \ref{fig:Total_particles} via the discrepancy of the lines for different $\chi$ values, and are much smaller than the associated error from $\Gamma_{c}$, indicated by the shaded region.

\end{document}